\documentclass[preprint,12pt]{elsarticle}

\usepackage{graphicx,subfigure}
\usepackage{amssymb}
\usepackage{amsmath}
\usepackage{bm}
\usepackage{lineno}
\usepackage{natbib}
\usepackage[squaren]{SIunits}

\begin{document}

\begin{frontmatter}

\title{Elimination of the numerical Cerenkov instability for spectral  EM-PIC codes}

\author[UCLAEE]{Peicheng Yu}
\author[THUACC]{Xinlu Xu}
\ead{xuxl10@mails.tsinghua.edu.cn}
\author[UCLAPH]{Viktor K. Decyk}
\author[LLNL]{Frederico Fiuza}
\author[IST]{Jorge Vieira}
\author[UCLAPH]{Frank S. Tsung}
\author[IST,ISCTE]{Ricardo A. Fonseca}
\author[THUACC]{Wei Lu}
\author[IST]{Luis O. Silva}
\author[UCLAEE,UCLAPH]{Warren B. Mori}

\address[UCLAEE]{Department of Electrical Engineering, University of California Los Angeles, Los Angeles, CA 90095, USA}
\address[THUACC]{Department of Engineering Physics, Tsinghua University, Beijing 100084, China}
\address[UCLAPH]{Department of Physics and Astronomy, University of California Los Angeles, Los Angeles, CA 90095, USA}
\address[LLNL]{Lawrence Livermore National Laboratory, Livermore, California, USA}
\address[IST]{GOLP/Instituto de Plasma e Fus\~ao Nuclear, Instituto Superior T\'ecnico, Universidade de Lisboa, Lisbon, Portugal}
\address[ISCTE]{ISCTE - Instituto Universit\'ario de Lisboa, 1649--026, Lisbon, Portugal}

\begin{abstract}
When using an electromagnetic particle-in-cell (EM-PIC) code to simulate a relativistically drifting plasma, a violent numerical instability known as the numerical Cerenkov instability (NCI) occurs. The NCI is due to the unphysical coupling of electromagnetic waves on a grid to wave-particle resonances, including aliased resonances, i.e., $\omega + 2\pi\mu/\Delta t=(k_1+ 2\pi\nu_1/\Delta x_1)v_0$, where $\mu$ and $\nu_1$ refer to the time and space aliases and the plasma is drifting relativistically at velocity $v_0$ in the $\hat{1}$-direction. We extend our previous work [X. Xu, et. al., Comp. Phys. Comm. 184, 2503 (2013)] by recasting the numerical dispersion relation of a relativistically drifting plasma into a form which shows explicitly how the instability results from the coupling modes which are purely transverse electromagnetic (EM) modes and purely longitudinal modes in the rest frame of the plasma for each time and space aliasing. The dispersion relation for each  $\mu$ and $\nu_1$ is the product of the dispersion relation of these two modes set equal to a coupling term that vanishes in the continuous limit. The new form of the numerical dispersion relation provides an accurate method of systematically calculating the growth rate and location of the mode in the fundamental Brillouin zone for any Maxwell solver for each $\mu$ and $\nu_1$. We then focus on the spectral Maxwell solver and systematically discuss its NCI modes. We show that the second fastest growing NCI mode for the spectral solver corresponds to $\mu=\nu_1=0$, that it has a growth rate approximately one order of magnitude smaller than the fastest growing $\mu=0$ and $\nu_1=1$ mode, and that its location in the $\bm{k}$ space fundamental Brillouin zone is sensitive to the grid size and time step. Based on these studies, strategies to systematically eliminate the NCI modes for a spectral solver are developed. We apply these strategies to both relativistic collisionless shock and LWFA simulations, and demonstrate that high-fidelity multi-dimensional simulations of drifting plasmas can be carried out with a spectral Maxwell solver with no evidence of numerical Cerenkov instability.
\end{abstract}

\begin{keyword}
Particle-in-cell \sep plasma simulation \sep relativistic drifting plasma  \sep numerical Cerenkov instability  \sep numerical dispersion relation\sep spectral solver
\end{keyword}

\end{frontmatter}


\section{Introduction}
The study of the multi-dimensional numerical Cerenkov instability (NCI) in electromagnetic particle-in-cell (EM-PIC) plasma simulations has attracted much renewed attention since the identification of this numerical instability as the limiting factor of Lorentz boosted frame simulations for laser wakefield acceleration  (LWFA) \cite{Vay2007PRL,Martins2010NatPhys,VayPoP,MartinsCPC}. Furthermore, the NCI is also a limiting factor in relativistic collisionless shock simulations \cite{shock1,shock2}. Past and recent work has shown that the NCI inevitably arises in EM-PIC simulations when a plasma (neutral or pure electron) drifts across a simulation grid with a speed near the speed of light. Analysis shows that it is due to the unphysical coupling of electromagnetic like modes and wave particle resonances (including those due to aliasing) \cite{Godfrey1974,Godfrey1975,YuAAC,GodfreyJCP2013,XuArxiv}. This instability leads to unphysical exponential growth of the EM field energy which interferes with the  physics being studied in the simulation. As a result, significant recent effort has been devoted to the understanding and elimination of the NCI so that high fidelity relativistic plasma drift simulations can be routinely performed \cite{VayJCP2011,YuAAC,GodfreyJCP2013,XuArxiv,YuArxiv,GodfreyPSATD,GodfreyFDTDnotes}.

In this recent work, dispersion relations for the NCI has been derived and analyzed. We begin by restating the dispersion tensor provided in Ref. \cite{XuArxiv}. This tensor can be used to study the pattern and growth rates of the NCI in Fourier space. For a cold plasma drifting in $\hat 1$-direction relativistically with the unperturbed normalized distribution function of
\begin{align}\label{eq:initdis}
f^n_0=\delta(p_1-p_0)\delta(p_2)\delta(p_3)
\end{align}
where $p_0=\gamma v_0$, and $v_0$ is the drifting velocity of the plasma, the corresponding dispersion relation for the plasma drift is \cite{XuArxiv}
 \begin{align}\label{eqdetepsilon}
\textrm{Det}(\overleftrightarrow{\epsilon})=0
\end{align}
where the elements of $\overleftrightarrow{\epsilon}$ are
\begin{align}\label{eqepsilon3d}
\epsilon_{11}&=[\omega]^2-[k]_{E2}[k]_{B2}-[k]_{E3}[k]_{B3}\nonumber\\
&-\frac{\omega^2_p}{\gamma}\sum_{\mu,\bm{\nu}}(-1)^{\mu}\frac{S_{j1}\{S_{E1}[\omega]\omega' / \gamma^2+v^2_0(S_{B3}k'_2[k]_{E2}+S_{B2}k'_3[k]_{E3})\}}{(\omega'-k'_1v_0)^2}\nonumber\\
\epsilon_{12}&= [k]_{E1}[k]_{B2} - \frac{\omega^2_p}{\gamma}\sum_{\mu,\bm{\nu}}(-1)^{\mu}\frac{S_{j1}v_0k'_2(S_{E2}[\omega]-v_0S_{B3}[k]_{E1})}{(\omega'-k'_1v_0)^2}\nonumber\\
\epsilon_{13}&=[k]_{E1}[k]_{B3}- \frac{\omega^2_p}{\gamma}\sum_{\mu,\bm{\nu}}(-1)^{\mu}\frac{S_{j1}v_0k'_3(S_{E3}[\omega]-v_0S_{B2}[k]_{E1})}{(\omega'-k'_1v_0)^2}\nonumber\\
\epsilon_{21}&=[k]_{E2}[k]_{B1}-\frac{\omega^2_p}{\gamma}\sum_{\mu,\bm{\nu}}(-1)^{\mu}\frac{v_0S_{j2}S_{B3}[k]_{E2}}{\omega'-k'_1v_0}\nonumber\\
\epsilon_{22}&=[\omega]^2-[k]_{E1}[k]_{B1}-[k]_{E3}[k]_{B3}-\frac{\omega^2_p}{\gamma}\sum_{\mu,\bm{\nu}}(-1)^{\mu}\frac{S_{j2}(S_{E2}[\omega]-v_0S_{B3}[k]_{E1})}{\omega'-k'_1v_0}\nonumber\\
\epsilon_{23}&=[k]_{E2}[k]_{B3}\nonumber\\
\epsilon_{31}&=[k]_{E3}[k]_{B1}-\frac{\omega^2_p}{\gamma}\sum_{\mu,\bm{\nu}}(-1)^{\mu}\frac{v_0S_{j3}S_{B2}[k]_{E3}}{\omega'-k'_1v_0}\nonumber\\
\epsilon_{32}&=[k]_{E3}[k]_{B2}\nonumber\\
\epsilon_{33}&=[\omega]^2-[k]_{E1}[k]_{B1}-[k]_{E2}[k]_{B2}-\frac{\omega^2_p}{\gamma}\sum_{\mu,\bm{\nu}}(-1)^{\mu}\frac{S_{j3}(S_{E3}[\omega]-v_0S_{B2}[k]_{E1})}{\omega'-k'_1v_0}
\end{align}
where $[\omega]$, $[\bm k]_{E,B}$ are the finite difference operators of the Maxwell Equation solver, $\bm S_j$, $\bm S_{E}$ and $\bm S_{B}$ are the current and field interpolation functions (the definitions for $[\omega]$, $[\bm k]_{E,B}$, $\bm S_j$, $\bm S_{E}$, and $\bm S_{B}$ for each Maxwell solver are given in the Appendix of Ref. \cite{XuArxiv}), $v_0$ is normalized by the speed of light, and $\omega_p$ is defined as
\begin{align}\label{eq:omegap}
\omega^2_p=\frac{4\pi q^2n_p}{m}
\end{align}
where $q$ and $m$ are the electron charge and rest mass,  $n_p$ the plasma density in the drifting frame, and
\begin{align}\label{eqkgwg}
\omega'&=\omega+\mu\omega_g\qquad \omega_g=\frac{2\pi}{\Delta t}\qquad \mu = 0,\pm 1, \pm 2, \ldots\nonumber\\
{k}'_i&={k}_i+\nu_i{k}_{gi}\qquad {k}_{gi}=\frac{2\pi}{\Delta {x}_i}\qquad \nu_i = 0,\pm 1, \pm 2, \ldots
\end{align}
Due to the use of finite space and time steps, these dispersion relations not only contain terms from the lowest order Brillouin zones ($\mu=0$ and $\bm\nu=\bm 0$), but also the space aliasing (summation over $\bm\nu$) and time aliasing (summation over $\mu$) terms \cite{Lindman1970Langdon1970,birdsalllangdon}. The elements of the interpolation functions $\bm{S}_{j,E,B}$, and finite difference operators $[\cdot]$ vary depending on the field solver (e.g., spectral \cite{dawsonrmp,lin}, Yee \cite{Yee}, or Karkkainen solver \cite{kark}), particle shape, current deposition algorithm, and field interpolation scheme (e.g., momentum or energy conserving).

The NCI modes can be found by solving Eq. (\ref{eqdetepsilon}) numerically or analytically by specifying a $\mu$ and $\bm \nu$ \cite{XuArxiv}. The unstable modes are found near the intersections of the beam resonances ($\omega' - k'_1v_0=0$), and the EM modes ($[\omega]^2=[k]^2$ in vacuum). Since the numerical dispersion relation of the EM modes is mainly determined by the Maxwell field solver used in the simulation, the location of the unstable NCI modes can be manipulated through the choice of the Maxwell solver.  For example, in \cite{XuArxiv} it was shown that using a spectral Maxwell solver, i.e., solving Maxwell's equations in $\bm k$-space  \cite{dawsonrmp,lin}, moves the fastest growing modes to a region in $\bm k$-space far away from the modes of physical interest. In this case, the fastest growing modes all come from the first spatial aliasing beam ($\mu=0$, $\nu_1=\pm 1$) resonance. For a Yee solver, the fast growing modes come from this resonance, as well as from the fundamental mode ($\mu=\nu_1=0$) which leads to growth at $\bm k$ that resides in the middle of the interesting physics. In \cite{XuArxiv} it was shown that the fastest growing NCI mode for a spectral solver could be eliminated by applying a low pass filter in the solver.

Based on these results the feasibility of using an EM-PIC code with spectral solver to eliminate the NCI for LWFA simulations in a Lorentz boosted frame was subsequently investigated in \cite{YuArxiv}. For weakly nonlinear laser drivers the agreement between the rest and Lorentz boosted frames was excellent for arbitrary $\gamma$. For nonlinear cases, the lineouts of the wakefield in Fig. 8 in Ref. \cite{YuArxiv} showed differences at the highest $\gamma$.  Based on these differences, in this paper we have examined further both the growth rates and methods for eliminating lower growing NCI unstable modes with an emphasis for the spectral solver. We note that Godfrey and co-workers have proposed methods to reduce the growth rate of the NCI for finite difference solvers \cite{GodfreyFDTDnotes}. They have also discussed the NCI characters of an FFT based analytical time domain solver and concentrate on time steps larger than the Courant limit of the spectral Maxwell solver \cite{GodfreyPSATD}. The work presented here can be viewed as complementary since there are advantages and disadvantages for each type of solver and for using large or small time steps.

To systematically investigate these additional unstable modes, we rewrite the numerical dispersion relation into the form of two coupled modes whose coupling term vanishes in the continuous limit with $\Delta t\rightarrow 0$ and $\Delta x_i\rightarrow 0$. These two coupled modes can be identified as the numerical form of the Lorentz transformation of modes which are purely longitudinal (plasma oscillations) and purely transverse (EM waves) in the rest frame of the plasma. These modes are uncoupled in the rest frame in the continuous limit. Therefore, in the continuous limit where Lorentz invariance is strictly true they are also decoupled, although no longer remaining purely longitudinal and transverse. However, we show that in the discrete limit there is a non-vanishing coupling term in the dispersion relation. The coupling term explicitly shows how the finite grid sizes and time step leads to the inevitable coupling between these modes which leads to instability.

Recasting the dispersion relation in this new form not only sheds light on the mechanism of the NCI, but also provides a natural way to systematically calculate the unstable NCI modes. We use this dispersion relation to obtain analytical expressions for the family of unstable modes which includes the dependence of growth rate on the grid size, time step, and plasma density; as well as the location of the modes in $\bm k$-space. We find excellent agreement between the analytical expressions, numerical solutions to the dispersion relation, and PIC simulations.

Based on these new results, we have experimented with different strategies for eliminating the first, second, and higher order unstable NCI modes with an emphasis on the spectral solver. As has been demonstrated in \cite{XuArxiv,YuArxiv}, when a low pass filter is used the fastest growing NCI modes [at $(\mu,\nu_1)=(0,\pm 1)$]  are completely eliminated in the spectral solver. In addition,  we find that for the parameter space of interest the second fastest growing modes are those at $(\mu,\nu_1)=(0,0)$. These modes have a highly localized pattern of four dots in Fourier space (one in each quadrant) in 2D and two rings in 3D. Since in some cases the unstable modes can reside near or in the middle of modes of physical interest, filtering them out directly could potentially affect the accuracy of the physics model. However, we find that unlike the fastest growing modes, the location and growth rate of these $(\mu,\nu_1)=(0,0)$ mode depend on the time step as well as the plasma density. As the time step is reduced these modes move to higher values of wave number well outside the region of physical interest and their growth rate is greatly reduced. Thus, these modes can be eliminated by reducing the time step. In addition, we find that by slightly modifying the finite difference operator $[k]_1$ of the Maxwell solver at the region where the $(\mu,\nu_1)=(0,0)$ modes reside, we can completely eliminate these modes. We also find that the location, growth rate, and the values of $(\mu,\nu_1)$ of the next fastest growing modes depend on the time step and particle shape, and reducing the time step can also effectively eliminate these modes as well. These results reveal clear advantages of spectral EM-PIC codes: the results always converge as the time step is reduced, which allows one to check the validity of any simulation.

We then present results from relativistic collisionless shock simulations, as well as LWFA simulations in the nonlinear regime using this new understanding of the location and growth rate of the unstable modes. In all these simulations involving relativistically drifting plasma we used a low-pass filter to eliminate the fastest growing NCI modes. We conducted simulations at time steps near the Courant limit, as well as at smaller time steps in which the second fastest growing modes have a smaller growth rate and reside farther away from the physical modes. We likewise used the method of modifying the EM dispersion curve to complete eliminate the $(\mu,\nu_1)=(0,0)$ modes. Comparing the results from these two approaches allow one to identify how the NCI and/or time step affect the physics modeling. More accurate modeling are observed in both shock simulations, and LWFA simulations when the elimination strategies are used. In the last section, we summarize the results and discuss areas for future work.

\section{Spectral solver NCI mode in 2D}
\label{sect:2d}
Without loss of generality, we systematically investigate the NCI modes by starting from the numerical dispersion relation of a 2D drifting plasma with the appropriate elements of the dispersion tensor $\overleftrightarrow{\epsilon}$ \cite{XuArxiv}
\begin{align}\label{eqepsilon2d}
\epsilon_{11}&= {[\omega]^2} -[k]_{E2}[k]_{B2}-{\frac{\omega^2_p}{\gamma}}\sum_{\mu,\bm\nu} (-1)^\mu \frac { S_{j1}(S_{E1}\omega' [\omega]/\gamma^2+ S_{B3}[k]_{E2}k_2'v_0^2)} {(\omega ' - k_1' v_0)^2}\nonumber\\
\epsilon_{12} &=[k]_{E1}[k]_{B2}-{\frac{\omega^2_p}{\gamma}}\sum_{\mu,\bm\nu} (-1)^\mu\frac{k_2'v_0S_{j1}(S_{E2}[\omega]- S_{B3}v_0[k]_{E1})}{(\omega'-k_1'v_0)^2}\nonumber\\
\epsilon_{21} &= [k]_{E2}[k]_{B1}-{\frac{\omega^2_p}{\gamma}}\sum_{\mu,\bm\nu} (-1)^\mu\frac{ S_{j2}S_{B3}[k]_{E2}v_0}{\omega'-k_1'v_0}\nonumber\\
\epsilon_{22}&= {[\omega]^2} -[k]_{E1}[k]_{B1}-{\frac{\omega^2_p}{\gamma}} \sum_{\mu,\bm\nu} (-1)^\mu\frac{S_{j2}(S_{E2}[\omega]- S_{B3}[k]_{E1}v_0) }{\omega' - k_1'v_0}\nonumber\\
\epsilon_{33}&={[\omega]^2} -[k]_{E1}[k]_{B1}-[k]_{E2}[k]_{B2}-{\frac{\omega^2_p}{\gamma}} \sum_{\mu,\bm\nu} (-1)^\mu\frac{S_{j3}(S_{E3}[\omega]- S_{B2}[k]_{E1}v_0)}{\omega' - k_1'v_0}\nonumber\\
\epsilon_{13}&=\epsilon_{23}=\epsilon_{31}=\epsilon_{32}=0
\end{align}
We focus on the $\bm k$ near the beam resonance line \cite{XuArxiv}
\begin{align}
\omega' - k'_1v_0=0.
\end{align}
The numerical solution of Eq. (\ref{eqdetepsilon})--(\ref{eqepsilon3d}) for each mode can be analytically obtained by keeping only the corresponding $\mu$ and $\bm\nu$ terms in Eq. (\ref{eqepsilon2d}) since these terms are dominant near the corresponding resonance lines. Note for the cases considered in this paper, we find it is a good approximation to truncate the sum of $\nu_2$ and only keep the $\nu_2=0$ term. The corresponding dispersion relation $\epsilon_{11}\epsilon_{22}-\epsilon_{12}\epsilon_{21}=0$ becomes
\begin{align}\label{eq:epsilon2deqzero}
&\left( [\omega]^2 - [k]_{E2}[k]_{B2} - \frac{\omega^2_p}{\gamma}(-1)^\mu \frac{ S_{j1} (S_{E1} [\omega]\omega' / \gamma^2 + S_{B3}v_0^2 [k]_{E2}k_2)}{(\omega'-k'_1v_0)^2}\right)\times\nonumber\\
& \left( [\omega]^2- [k]_{E1}[k]_{B1} - \frac{\omega^2_p}{\gamma}(-1)^\mu\frac{S_{j2} (S_{E2}[\omega] - S_{B3} [k]_{E1} v_0)}{\omega' - k'_1 v_0} \right) - \nonumber \\
&  \left([k]_{E1} [k]_{B2} - \frac{\omega^2_p}{\gamma}(-1)^\mu\frac{S_{j1} v_0k_2 (S_{E2} [\omega] - S_{B3}[k]_{E1} v_0)}{ (\omega'-k'_1v_0)^2} \right)\times\nonumber\\
&\left( [k]_{E2} [k]_{B1} - \frac{\omega^2_p}{\gamma}(-1)^\mu\frac{S_{j2} S_{B3} v_0[k]_{E2} }{\omega' - k'_1v_0} \right)=0
\end{align}
After some algebra, Eq. (\ref{eq:epsilon2deqzero}) can be written as
\begin{align}\label{eq:2dmodesres}
& \left((\omega' - k'_1 v_0)^2- \frac{\omega_p^2}{\gamma^3}(-1)^\mu\frac{S_{j1} S_{E1}\omega'}{[\omega]} \right)\times  \nonumber\\
&\left( [\omega]^2 - [k]_{E1}[k]_{B1}  - [k]_{E2}[k]_{B2}  - \frac{\omega_p^2}{\gamma} (-1)^\mu\frac{S_{j2}(S_{E2}[\omega] - S_{B3} [k]_{E1} v_0)}{\omega' - k'_1 v_0} \right) \nonumber\\
&+ \mathcal{C}=0
\end{align}
where $\mathcal{C}$ is a coupling term in the dispersion relation
\begin{align}\label{eq:coupling}
\mathcal{C}=\frac{\omega_p^2}{\gamma} \frac{(-1)^\mu}{[\omega]}\biggl\{&S_{j1}S_{E1}\omega'[k]_{E2}[k]_{B2}(v^2_0-1)+S_{j2}S_{E2}[k]_{E2}[k]_{B2}(\omega'-k'_1v_0)\nonumber\\
&+S_{j1}[k]_{E2}(S_{E2}[k]_{B1}k_2v_0-S_{B3}k_2v^2_0[\omega])\biggr\}
\end{align}

Much can be learned by investigating Eq. (\ref{eq:2dmodesres}). First, in the continuous limit ($\Delta t\rightarrow 0$, $\Delta x_i\rightarrow 0$, and $\nu_1=0$), we have $[\omega]\rightarrow \omega$, $S_{E,B}\rightarrow 1$, so the coupling term $\mathcal{C}$ vanishes; second, the two factors in the first term of Eq. (\ref{eq:2dmodesres}) are the Lorentz transformation of the dispersion relation of the Langmuir (longitudinal) mode, and the EM (transverse) mode in a stationary plasma, which in the continuous limit reduce to
\begin{align}
\label{eq:langmuir}(\omega - k_1 v_0)^2- \frac{\omega_p^2}{\gamma^3} = 0\qquad \omega^2 - k_1^2  - k_2^2  - \frac{\omega_p^2}{\gamma} = 0
\end{align}
Consequently, we can identify the numerical Langmuir modes and EM modes for a drifting plasma as
\begin{align}
\label{eq:numlangmuir}(\omega' - k'_1 v_0)^2- \frac{\omega_p^2}{\gamma^3}(-1)^\mu\frac{S_{j1} S_{E1}\omega'}{[\omega]}  &\approx 0\\
\label{eq:em}  [\omega]^2 - [k]_{E1}[k]_{B1}  - [k]_{E2}[k]_{B2}  - \frac{\omega_p^2}{\gamma} (-1)^\mu\frac{S_{j2}(S_{E2}[\omega] - S_{B3} [k]_{E1} v_0)}{\omega' - k'_1 v_0}&\approx 0
\end{align}
In addition, from Eq. (\ref{eq:2dmodesres}) we see that when finite grid sizes and time steps are used neither Eq. (\ref{eq:numlangmuir}) nor Eq. (\ref{eq:em}) leads to instabilty (if the Courant condition is satisfied). Therefore it becomes clear that the NCI is caused by the numerical coupling between modes which are purely longitudinal and purely transverse in the plasma rest frame due to the non-vanishing term $\mathcal{C}$. In \cite{XuArxiv} the conclusion that NCI can be found near the intersections of the EM modes and Langmuir modes (or equivalently $\omega'-k'_1v_0=0$ since the $\omega^2_p/\gamma^3$ term is negligible) is obtained from examining the simulation and numerical data. With the new form Eq. (\ref{eq:2dmodesres}), we can now directly see how the Langmuir mode couples to the EM modes. Therefore, reducing or eliminating the coupling term $\mathcal{C}$ is the key to mitigating the NCI. Another interesting fact obtained from Eq. (\ref{eq:2dmodesres}) is that, if we assume that the  $\omega^2_p$ term in Eq. (\ref{eq:numlangmuir}) and (\ref{eq:em}) are small and can be neglected, when determining the positions of these two modes in Fourier space,  the time and space aliasing $\mu$ and $\nu_1$ are in the Langmuir modes, while there is no aliasing part in the EM mode. As a side note, it is evident from Eq. (19) of \cite{XuArxiv} that in 1D the coupling term vanishes, i.e. $\mathcal{C}=0$ in the numerical dispersion relation, hence no NCI is found in 1D.

\begin{figure}[th]
\begin{center}\includegraphics[width=.8\textwidth]{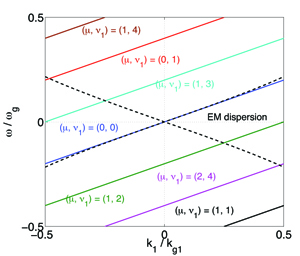}
\caption{\label{fig:beamresonance} The EM dispersion relation together with the beam resonance $\omega'-k'_1\beta=0$ is shown. The parameters used to plot this figure are listed in Table \ref{tab:simpara}.}
\end{center}
\end{figure}

For each pair of $(\mu,\nu_1)$ there is a corresponding Eq. (\ref{eq:2dmodesres}). However, in PIC algorithm the range of $(\omega, k_1)$ for the quantities defined at discrete locations and time step is limited to the fundamental Brillouin zone $k_i\in (-k_{gi}/2,k_{gi}/2)$, $\omega\in(-\omega_g/2,\omega_g/2)$. As a result, not all the $(\mu,\nu_1)$ wave-particle resonance line exist within the fundamental Brillouin zone. In the following, we describe a way to systematically identify  the wave-particle resonance lines inside the fundamental zone. Take the parameters in Table \ref{tab:simpara} as an example, we first plot the $(\mu,\nu_1)=(0,0)$ line [blue line in Fig. \ref{fig:beamresonance}]. As the line extends to the right it meets the boundary of the fundamental zone at $k_1=0.5k_{g1}$. To further extend it into the fundamental zone we add $\nu_1$ by 1, fold the line to the right boundary of $k_1$, and obtain the $(\mu,\nu_1)=(0,1)$ line [red line in Fig. \ref{fig:beamresonance}]. The red line extends further until it reaches the $\omega=0.5\omega_g$ boundary. To  extend it further  we increase $\mu$ by 1, and obtain the $(\mu,\nu_1)=(1,1)$ line. More higher order modes in the fundamental Brillouin zone can be obtained in this way. The negative $(\mu,\nu_1)$ lines can likewise be obtained by starting from the main Langmuir mode and then extending it to the left, and sets of these $(\mu,\nu_1)$ lines can be obtained as the lines hit the boundary at $\omega=-0.5\omega_g$ ($\mu$ is reduced by 1) and $k_1=-0.5k_{g1}$ ($\nu_1$ is reduced by 1). Using the normalization
\begin{align}
\hat\omega+\mu=v_0 (\hat{k}_1+\nu_1) \lambda_1
\end{align}
where
\begin{align}\label{eq:hatwk}
\hat{\omega}=\frac{\omega}{\omega_g}\qquad \hat{k}_i=\frac{k_i}{k_{gi}}\qquad \lambda_i = \frac{\Delta t}{\Delta x_i}
\end{align}
the criterion for the Langmuir modes to be inside the fundamental Brillouin zone are $\vert v_0 \lambda_1 \nu_1 - \mu \vert < 0.5 + 0.5 v_0\lambda_1$. 
Note for explicit Maxwell solvers $\lambda_1<1$ is a requirement for stable propagation of EM waves in vacuum. The NCI occurs where a resonance line intersects the EM dispersion relation. In Fig. \ref{fig:beamresonance} we also plot the EM dispersion relation in vacuum as dashed lines. Note for EM curves we only show $\hat{\omega}$ v.s. $\hat{k}_1$ at $\hat k_2=0$, but this line varies as $\hat{k}_2$ changes. For the NCI pattern and growth rates associated with each resonance line, we can numerically solve Eq. (\ref{eq:2dmodesres}) using the corresponding $\mu$ and $\nu_1$. Note in \cite{GodfreyJCP2013} a plot similar to Fig. \ref{fig:beamresonance} can be found (Fig. 1 of Ref. \cite{GodfreyJCP2013}). However, in \cite{GodfreyJCP2013} all the $\mu$ are summed over analytically, while in this paper we emphasize that for a particular resonance line, only one $\mu$ term in the elements of $\overleftrightarrow\epsilon$ is playing a dominant role. Furthermore, care should be taken when summing over $\mu$ and $\nu_1$ as they are not independent sums.

While Eq. (\ref{eq:2dmodesres}) can be used to study the fastest growing mode at $(\mu,\nu_1)=(0,\pm 1)$ which was investigated in Ref. \cite{XuArxiv} and \cite{YuArxiv}, here we concentrate on the additional modes. We use Eq. (\ref{eq:2dmodesres}) to develop analytical expressions within the parameter space we are interested in. Starting from Eq. (\ref{eq:2dmodesres}), we expand $\omega'$ around the beam resonance $\omega' = k'_1v_0$, and write $\omega' = k'_1v_0 + \delta\omega'$, where $\delta\omega'$ is a small term. In addition, we use the relativistic limit $v_0\rightarrow 1$, and expand the finite difference operator $[\omega]$ as
\begin{align}\label{eq:omegafirstorder}
[\omega]\approx [\omega]\biggl\vert_{\tilde k_1v_0}+\delta\omega'\frac{\partial [\omega]}{\partial \omega}\biggl\vert_{\tilde k_1v_0}\qquad\qquad
\end{align}
where
\begin{align}
[\omega]\biggl\vert_{\tilde k_1v_0}\equiv\xi_0=\frac{\sin(\tilde k_1\Delta t/2)}{\Delta t/2}\qquad \frac{\partial [\omega]}{\partial \omega}\biggl\vert_{\tilde k_1v_0}\equiv\xi_1 = \cos(\tilde k_1\Delta t/2)
\end{align}
where $\tilde k_1=k_1+\nu_1 k_{g1}-\mu\omega_g$, and $[\omega]^2\approx \xi^2_0+2\xi_0\xi_1\delta\omega'$. In addition, we found it is sufficiently accurate if we neglect the $\omega^2/\gamma^3$ term in the Langmuir mode in Eq. (\ref{eq:2dmodesres}). This is why it is essentially the same to say that the instability occurs at wave-particle resonances, beam resonances, or at Langmuir resonances. Moreover, note that $\omega$ terms likewise appear in $\bm S_B$ (see Appendix of \cite{XuArxiv}), and we will separate it from $\bm S_B$ by writing
\begin{align}
\bm S_B=\cos(\omega\Delta t/2)\bm S'_B
\end{align}
and expand $\bm S_B$ to first order as
\begin{align}
\bm S_B=(\zeta_0+\zeta_1\delta\omega')\bm S'_B\qquad \zeta_0\equiv \cos(\tilde k_1\Delta t/2) \qquad \zeta_1\equiv -\sin(\tilde k_1\Delta t/2)\Delta t/2
\end{align}
Using these approximations, we obtain a cubic equation for $\delta\omega'$,
\begin{align}\label{eq:asym1}
A_2\delta\omega'^3+B_2\delta\omega'^2+C_2\delta\omega'+D_2=0
\end{align}
where
\begin{align}
A_2=&2\xi^3_0\xi_1\nonumber\\
B_2=&\xi^2_0\biggl\{\xi^2_0-[k]_{E1}[k]_{B1}-[k]_{E2}[k]_{B2}-\frac{\omega^2_p}{\gamma}(-1)^\mu S_{j2}(S_{E2}\xi_1-\zeta_1S'_{B3}[k]_{E1})\biggr\}\nonumber\\
C_2=&\frac{\omega^2_p}{\gamma}(-1)^\mu \biggl\{ \xi^2_0S_{j2}(\zeta_0S'_{B3}[k]_{E1}-S_{E2}\xi_0)-{\xi_1}S_{j1}S_{E2}[k]_{E2}k_2[k]_{B1}\nonumber\\
& +\xi_0[k]_{E2}(S_{j2}S_{E2}[k]_{B2}-S_{j1}k_2\zeta_1S'_{B3}\xi_0)\biggr\}\nonumber\\
\label{eq:asym2gen}D_2=&\frac{\omega^2_p}{\gamma}(-1)^\mu \xi_0[k]_{E2}k_2S_{j1}\biggl( S_{E2}[k]_{B1}-\zeta_0S'_{B3}\xi_0\biggr)
\end{align}
The coefficients $A_2$ to $D_2$ are real, and completely determined by $k_1$ and $k_2$. When the discriminant of this cubic equation
\begin{align}\label{eq:discriminat}
\Delta = 18A_2B_2C_2D_2-4B^3_2D_2+B^2_2C^2_2-4A_2C_2-27A^2_2D^2_2
\end{align}
satisfies the condition $\Delta < 0$, the cubic equation has one real root and two non-real complex conjugate roots. Therefore, by calculating the discriminant of the cubic equation Eq. (\ref{eq:discriminat}), we can quickly identify the position of the instability for a particular $\nu_1$. We can then use the general formula for the roots of a cubic equation to obtain the growth rate of the corresponding $\bm k$ mode. As a result, by solving Eqs. (\ref{eq:asym1}) and (\ref{eq:asym2gen}) we can rapidly calculate the location and growth rate of the instability.

Before we discuss the predictions of Eqs. (\ref{eq:asym1}) and (\ref{eq:asym2gen}), we discuss how these equations differ from those used in our earlier work. In Ref. \cite{XuArxiv}, we derived a cubic equation to calculate the $(\mu,\nu_1)=(0,\pm 1)$ and $(\mu,\nu_1)=(0,0)$ modes of the Yee solver and Karkkainen solver, as well as the $(\mu,\nu_1)=(0,\pm 1)$ modes of the spectral solver [see Eqs. (33) and (34) in \cite{XuArxiv}, and Eq. (\ref{eq:coefold}) below]. While the expression works well for the modes we studied in \cite{XuArxiv}, it does not work well for the $(\mu,\nu_1)=(0,0)$ modes in the spectral solver which are the main focus of this paper. In \cite{XuArxiv} the coefficient of the cubic equation are
\begin{align}\label{eq:coefold}
A_2&=2\xi^3_0\xi_1\nonumber\\
B_2&=\xi^2_0(\xi^2_0-[k]_{E1}[k]_{B1}-[k]_{E2}[k]_{B2})\nonumber\\
C_2&=[k]_{E1}[k]_{B2}Q_{21}-(\xi^2_0-[k]_{E2}[k]_{B2})Q_{22}\nonumber\\
D_2&=-(\xi^2_0-[k]_{E1}[k]_{B1})Q_{11}+[k]_{E2}[k]_{B1}Q_{12}
\end{align}
where
\begin{align}
Q_{11}&=\frac{\omega^2_p}{\gamma}S_{j1}(S_{E1}\omega'[\omega]/\gamma^2+S_{B3}[k]_{E2}k_2)\qquad Q_{12}=\frac{\omega^2_p}{\gamma}k_2S_{j1}(S_{E2}[\omega]-S_{B3}[k]_{E1})\nonumber\\
Q_{21}&=\frac{\omega^2_p}{\gamma}S_{j2}S_{B3}[k]_{E2}\qquad Q_{22}=\frac{\omega^2_p}{\gamma}S_{j2}(S_{E2}[\omega]-S_{B3}[k]_{E1})\nonumber\\
Q_{33}&=\frac{\omega^2_p}{\gamma}S_{j3}(S_{E3}[\omega]-S_{B2}[k]_{E1})\qquad Q_{13}=Q_{23}=Q_{31}=Q_{32}=0
\end{align}
in 2D. In \cite{XuArxiv} it was implicitly assumed that the coefficients, $Q_{ij}$ were only expanded to zeroth order in powers of $\delta\omega'$. Otherwise, the coefficients $C_2$ and $D_2$ would themselves depend on $\delta \omega'$. This assumption was based on the fact that each $Q_{ij}$ is proportional to $\omega_p^2/\gamma$ which is itself small. We used the resulting cubic equation for $\delta\omega'$ to rapidly scan for unstable NCI modes in the fundamental Brillouin zone in $(k_1,k_2)$ space. Essentially within the $Q_{ij}$ coefficients we set $[\omega]\approx \xi_0$ and $\cos(\omega\Delta t/2)\approx \zeta_0$.
The corresponding coefficients for the cubic equation for $\delta\omega'$ then became
\begin{align}
A_2=&2\xi^3_0\xi_1\nonumber\\
B_2=&\xi^2_0\biggl\{\xi^2_0-[k]_{E1}[k]_{B1}-[k]_{E2}[k]_{B2}\biggr\}\nonumber\\
C_2=&\frac{\omega^2_p}{\gamma}(-1)^\mu \biggl\{ \xi^2_0S_{j2}(\zeta_0S'_{B3}[k]_{E1}-S_{E2}\xi_0)+\xi_0[k]_{E2}(S_{j2}S_{E2}[k]_{B2}-S_{j1}\zeta_0S'_{B3}k_2\xi_1)\biggr\}\nonumber\\
D_2=&\frac{\omega^2_p}{\gamma}(-1)^\mu \xi_0 [k]_{E2}k_2S_{j1}\biggl( S_{E2}[k]_{B1}-\zeta_0S'_{B3}\xi_0\biggr)
\end{align}
The resulting expression was successfully used to examine the $(\mu,\nu_1)=(0,\pm 1)$ modes of the Yee, Karkkainen, and spectral solver, as well as the $(\mu,\nu_1)=(0,\pm 1)$ modes of the Yee, and Karkkainen solver. However, it did not predict the unstable $(\mu,\nu_1)=(0,0)$ modes for the spectral solver.

We now return to Eqs. (\ref{eq:2dmodesres}), (\ref{eq:asym1}), and (\ref{eq:asym2gen}) with an eye toward the spectral solver which is the main focus of this paper. For this case $[\bm k]_{E,B}=\bm k$, therefore in 2D
\begin{align}
S_{E1}&=S_{E2}=S_{E3}\equiv S_E=S_l\qquad S_{B1}=S_{B2}=S_{B3}\equiv S_B=\cos\frac{ \omega\Delta t}{2}S_l.\nonumber\\
S_{j1}&=S_{j2}=S_{j3}\equiv S_E=S_l\nonumber
\end{align}
where
\begin{align}
S_l=\biggl(\frac{\sin(k_1\Delta x_1/2)}{k_1\Delta x_1/2}\biggr)^{l+1}\biggl(\frac{\sin(k_2\Delta x_2/2)}{k_1\Delta x_2/2}\biggr)^{l+1}
\end{align}
and $l$ corresponds to the order of the particle shape. Eq. (\ref{eq:2dmodesres}) reduces to
\begin{align}\label{eq:2dmodesresspectral}
&\left((\omega' - k'_1 v_0)^2- \frac{\omega_p^2}{\gamma^3}(-1)^\mu\frac{S_j S_E\omega'}{[\omega]} \right)\left( [\omega]^2 - k_1^2  - k_2^2  - \frac{\omega_p^2}{\gamma}(-1)^\mu S_j \frac{S_E[\omega] - S_B k_1 v_0}{\omega' - k'_1 v_0} \right)  \nonumber \\
&+ \frac{\omega_p^2}{\gamma[\omega] }(-1)^\mu S_jk_2^2 \{v_0^2(S_E\omega' - S_B[\omega])-v_0\nu_1S_Ek_{g1}\}=0
\end{align}
And the coefficients $A_2$ to $D_2$ of Eq. (\ref{eq:asym2gen}) becomes
\begin{align}
A_2&=2\xi^3_0\xi_1\nonumber\\
B_2&=\xi^2_0\biggl\{\xi^2_0-k^2_1-k^2_2-\frac{\omega^2_p}{\gamma}(-1)^\mu S^2_l(\xi_1-\zeta_1k_1)\biggl\}\nonumber\\
C_2&=\frac{\omega^2_p}{\gamma}(-1)^\mu S^2_l\biggl\{\xi^2_0(\zeta_0k_1-\xi_0)+\xi_0k_2(k_2-\zeta_0k_2\xi_1-k_2\zeta_1\xi_0)-\xi_1k_2^2(k_1-\zeta_0\xi_0)\biggl\}\nonumber\\
\label{eq:asym2}D_2&=\frac{\omega^2_p}{\gamma}(-1)^\mu \xi_0 S^2_lk_2^2(k_1-\zeta_0\xi_0)
\end{align}
The coefficients are real and completely determined by $k_1$ and $k_2$. We note that this dispersion relation could have been obtained from Eq. (23) if more terms were kept in the expansions for the $Q_{ij}$.

\begin{figure}[th]
\begin{center}\includegraphics[width=1\textwidth]{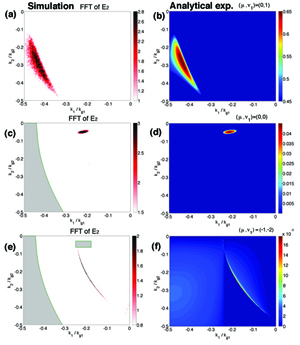}
\caption{\label{fig:variousnu1} (a), (c), and (e) are the FFT of $E_2$ in the 2D simulations using the parameters listed in Table \ref{tab:simpara}. The filter applied in order to observe these modes are illustrated by the grey areas in the plots. (b), (d), and (f) are the corresponding predictions by using the expression Eq. (\ref{eq:asym1}) and (\ref{eq:asym2}).}
\end{center}
\end{figure}

We now use the cubic equation for $\delta\omega'$ for the coefficients in Eq. (28) to systematically investigate the NCI modes for the spectral solver. In Fig. \ref{fig:variousnu1} (b), (d), and (f) we present the three sets of modes with the highest growth rate calculated by the analytical expressions Eqs. (\ref{eq:asym1}) and (\ref{eq:asym2}), for the parameters listed in Table \ref{tab:simpara}, and for linear particle shapes ($l=1$). Fig. \ref{fig:variousnu1} (b) shows the modes with $(\mu,\nu_1)=(0,\pm 1)$, which are the fastest growing NCI modes. These modes were already studied in \cite{XuArxiv}. Fig. \ref{fig:variousnu1} (d) shows the $(\mu,\nu_1)=(0,0)$ modes, which have a highly localized pattern of four dots [note that in (d) only one quadrant is plotted]. These modes usually have a maximum growth rate one order of magnitude smaller than the $(\mu,\nu_1)=(0,\pm 1)$ modes. For the parameters listed in Table \ref{tab:simpara}, the next fastest growing modes are the $(\mu,\nu_1)=(\pm 1,\pm 2)$ modes which have a maximum growth rate approximately 3 times smaller than the $(\mu,\nu_1)=(0,0)$ modes (for linear particle shape).

We have similarly performed UPIC-EMMA simulations in 2D to observe various NCI modes in the spectral solver, and to compare with the theory presented above. The simulations use a neutral plasma drifting at relativistic velocity, with the Lorentz factor $\gamma=50.0$. The plasma has a uniform initial spatial distribution, and we used the parameters listed in Table \ref{tab:simpara}. Note these parameters are those commonly used in the LWFA simulation in the Lorentz boosted frame \cite{YuArxiv}, and the plasma density is 100 times larger than that used in \cite{XuArxiv}.

\begin{table}[t]
\centering
\begin{tabular}{p{8cm}c}
\hline\hline
\textbf{Parameters} & \textbf{Values}\\
\hline
grid size $(k_0\Delta x_1, k_0\Delta x_2)$ & $(0.2,0.2)$\\
time step $\omega_0\Delta t$ & $0.4\Delta x_1$\\
boundary condition & Periodic \\
simulation box size $(k_0L_1,k_0L_2)$ & 102.4$\times 102.4$\\
plasma drifting Lorentz factor & $\gamma=50.0$\\
plasma density & $n_p/n_0 = 100.0$\\
\hline\hline
\end{tabular}
\caption{Crucial simulation parameters for the 2D relativistic plasma drift simulation. $n_0$ is the reference density, and $\omega^2_0=4\pi q^2n_0/m_e$, $k_0=\omega_0$ ($c$ is normalized to 1). }
\label{tab:simpara}
\end{table}

Fig. \ref{fig:variousnu1} (a), (c), and (e) show the simulation data of the FFT of $E_2$ at a particular time during the exponential EM energy growth from the NCI \cite{XuArxiv}. Fig. \ref{fig:variousnu1} (a) shows results from a simulation with no low-pass filter, and the most prominent modes are the $(\mu,\nu_1)=(0,\pm 1)$ modes that were analyzed in detail  in Ref. \cite{XuArxiv}. To generate the frames in the middle row, we use a low-pass filter to eliminate the $(\mu,\nu_1)=(0,\pm 1)$ modes. This makes the unstable $(\mu,\nu_1)=(0,0)$ modes more noticeable. It is shown in Fig. \ref{fig:variousnu1} (c) that the $(\mu,\nu_1)=(0,0)$ modes have a highly localized pattern of four dots [in Fig. \ref{fig:variousnu1} only one quadrant is shown], which agrees with the prediction of the analytic expression. According to Fig. \ref{fig:beamresonance}, there is no intersection between  $(\mu,\nu_1)=(1,1)$ resonance [and $(\mu,\nu_1)=(-1,-1)$ resonance] and the EM dispersion relation, so the next set of modes of interest are the $(\mu,\nu_1)=( 1,2)$ and $(\mu,\nu_1)=(-1,-2)$ modes. To make the $(\mu,\nu_1)=(\pm 1,\pm 2)$ mode more noticeable, we use a low-pass filter to filter out the $(\mu,\nu_1)=(0,\pm 1)$ mode, plus a four-dot mask filter to remove the $(\mu,\nu_1)=(0,0)$ mode. As shown in Figs. \ref{fig:variousnu1} (e) and (f), the locations of these modes in the simulation agree with the analytic prediction. As a side note, this numerical experiment also shows the simplicity and flexibility of using filters (masks) with complicated shapes in a spectral EM-PIC code to control the unphysical NCI growth.

According to both the theory and simulations, in the parameter space we are interested in, we usually categorize the NCI for a spectral solver into three categories: the fastest growing modes at $(\mu,\nu_1)=(0,\pm 1)$; the second fastest growing modes at $(\mu,\nu_1)=(0,0)$; and higher order NCI modes with $\vert\nu_1\vert> 1$ that have an even smaller growth rate. In the following we will discuss how the locations and positions of these modes change with the simulation parameters.

For the NCI modes with $|\nu_1| \ge 1$, the instability resides around the intersections of the Langmuir mode and EM mode \cite{XuArxiv,YuArxiv} (taking the small time step limit):
\begin{align}
(1-v^2_0)k^2_1+k^2_2-2\beta\xi k_1-\xi^2=0
\end{align}
where $\xi=\beta\nu_1k_{g1}-\mu \omega_g$. If we use the normalization in Eq. (\ref{eq:hatwk}) the equations above can be written as  (for square cells)
\begin{align}\label{eq:nu1position}
(1-v^2_0)\hat k^2_1+\hat k^2_2-2\beta\hat\xi \hat k_1-\hat\xi^2=0
\end{align}
where $\hat\xi=\beta\nu_1-\mu /\lambda_1$. The positions of the unstable NCI modes in $\bm{\hat k}$ space depends only on $\Delta t$ and $\Delta x_i$ through their ratio $\lambda_i$. Therefore, the position of the $|\nu_1| \ge 1$ NCI does not change if one keeps the ratio of time step to cell size. Moreover, if $\mu=0$, $\lambda_1$ does not appear in Eq. (\ref{eq:nu1position}), which means that the position of the $(\mu,\nu_1)=(0,\pm 1)$ modes are not affected by the time step.

\begin{figure}[th]
\begin{center}\includegraphics[width=1\textwidth]{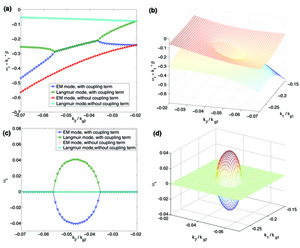}
\caption{\label{fig:rooteq} Roots of Eq. (\ref{eq:2dmodesres}) under the parameters listed in Table \ref{tab:simpara}. (a) and (c) shows the real, and imaginary parts of the roots between $\hat{k}_1=0.21$, and $-0.07\le \hat{k}_2 \le -0.2$, both with and without the coupling terms; meanwhile (b) and (d) shows the real and imaginary part of the roots in the range $-0.28\le \hat{k}_1\le -0.15$ and $-0.07\le \hat{k}_2 \le -0.02$.}
\end{center}
\end{figure}

For the NCI at $(\mu,\nu_1)=(0,0)$, there is no intersection between the corresponding fundamental Langmuir mode and the EM mode [as can be seen by plotting Eq. (\ref{eq:numlangmuir}) and (\ref{eq:em}) in $(\omega,k_1)$ space, see Fig. \ref{fig:rooteq} (a)]. The two modes interact at highly localized positions determined by the coupling term in Eq. (\ref{eq:2dmodesres}). To show how the coupling term in Eq. (\ref{eq:2dmodesres}) modifies the Langmuir and EM modes, we plot the solution of Eq. (\ref{eq:2dmodesres}) at $\hat{k}_1\approx 0.21$, $-0.07\le \hat{k}_2 \le -0.02$, where the instability is observed. Equation (\ref{eq:2dmodesres}) is solved both with, and without the coupling term (numerically forcing the coupling term to be zero). The parameters used in solving Eq. (\ref{eq:2dmodesres}) numerically are the same as in Table \ref{tab:simpara}, with $(\mu,\nu_1)=(0,0)$. It is evident in Fig. \ref{fig:rooteq} (a) and (c) that when the coupling term is present, the fundamental Langmuir mode and EM mode are coupled near $-0.057\le \hat{k}_2 \le -0.037$. In Fig. \ref{fig:rooteq} (c) where the growth rate is plotted, it becomes clear that in this range of $k_2$ where the fundamental Langmuir mode and EM mode are coupled, the two modes become complex conjugate pairs with one of them corresponding to instability in this range of $\hat k_2$. In  Figs. \ref{fig:rooteq} (b) and (d), we scan ranges in both $\hat k_1$ and $\hat k_2$, specifically, we scan  the range $\hat{k}_1\in[-0.28, -0.15]$ and $\hat{k}_2\in[-0.07,-0.02]$.

\begin{figure}[th]
\begin{center}\includegraphics[width=1\textwidth]{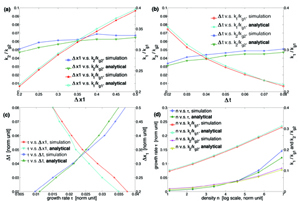}
\caption{\label{fig:tau_dt_dx_density} Dependence of the position $(\hat k_1,\hat k_2)$, as well as the growth rate $\tau$ of the NCI at the fundamental Langmuir mode to grid sizes $\Delta x_1$ (with $\Delta x_1=\Delta x_2$ fixed), time step $\Delta t$, and plasma density $n_p$. }
\end{center}
\end{figure}

We next investigate the sensitivity of the growth rate and location in $\bm k$-space to the simulation parameters for the NCI at the fundamental mode $(\mu,\nu_1)=(0,0)$. Note that we define the position of these modes at the value of $(\hat k_1,\hat k_2)$ where the growth rate is maximum.
In reality
there is a range (although highly localized) of modes that go unstable. Fig. \ref{fig:tau_dt_dx_density} (a)--(d) shows how the positions and growth rates of those modes change with plasma density and time step. For each simulation setup we plot both the simulation results and the predictions from the analytical expressions. When changing the grid sizes we fix $\Delta x_1=\Delta x_2$. Fig. \ref{fig:tau_dt_dx_density} (a) shows that when the grid sizes increases, the position of the $(\mu,\nu_1)=(0,0)$ NCI moves farther away from the center of the $(k_1,k_2)$ plot where the interesting real physics resides [red curve in Fig. \ref{fig:tau_dt_dx_density} (a)]. We keep $\Delta t$ constant as $\Delta x_1$ changes in Fig. \ref{fig:tau_dt_dx_density} (a). The  $(\mu,\nu_1)=(0,0)$ mode also moves farther away from the interesting physics when the time step decreases [see red curve in Fig. \ref{fig:tau_dt_dx_density} (b)]. Furthermore, as shown in Fig. \ref{fig:tau_dt_dx_density} (c), the growth rate decreases as the time step decreases [blue curve], which is not the case for the fastest growing modes of the NCI. The growth rate also decreases when the grid size increases while keeping $\Delta t$ fixed [Fig. \ref{fig:tau_dt_dx_density} (c) red curve]. When the density of the plasma increases (while fixing $\gamma_b=50$), the position of the $(\mu,\nu_1)=(0,0)$ NCI moves away from the center in $(k_1,k_2)$ space, and the growth rates of these modes increase [Fig. \ref{fig:tau_dt_dx_density} (d)].

\begin{figure}[th]
\begin{center}\includegraphics[width=1\textwidth]{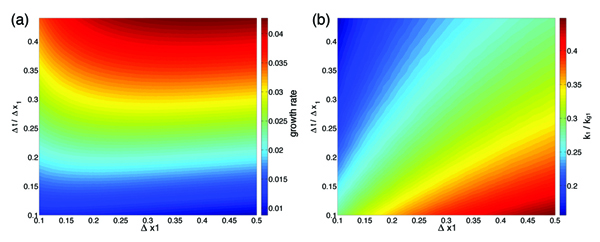}
\caption{\label{fig:analytical} Dependence of the (a) growth rate, and (b) $k_1$ position of NCI at the fundamental Langmuir mode for grid sizes $0.1 \le \Delta x_1\le 0.5$ (with $\Delta x_1=\Delta x_2$ fixed), and $0.1\le \Delta t/\Delta x_1 \le 0.45$.}
\end{center}
\end{figure}

\begin{figure}[th]
\begin{center}\includegraphics[width=1\textwidth]{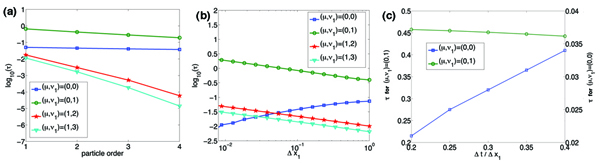}
\caption{\label{fig:allmodesgr}  (a) shows the dependence of the growth rate on particle shapes. (b) shows the dependence of the growth rate on grid size for various $\nu_1$ modes. (c) shows the dependence of the growth rate on time step when the grid sizes are fixed. Parameters listed in Table \ref{tab:simpara} are used for these plots.}
\end{center}
\end{figure}

A parameter scan which shows how the growth rate and position of the $(\mu,\nu_1)=(0,0)$ modes change with different choices of the grid sizes and time step, is shown in Fig. \ref{fig:analytical}. Note that we are keeping $\Delta x_1=\Delta x_2$ in the parameter scan. By examining Fig. \ref{fig:analytical}, we see that by reducing the $\Delta t/\Delta x_1$ ratio, the instability at the fundamental Langmuir mode moves farther towards larger $\hat k_1$ and the growth rate decreases. This is a unique characteristic of the $(\mu,\nu_1)=(0,0)$ modes, i.e., the growth rate of the fastest growing modes does not decrease as $\Delta t/\Delta x_1$ decreases. This is illustrated in Fig. \ref{fig:allmodesgr} (c) where the growth rate of the $(\mu,\nu_1)=(0,0)$ and $(0,1)$ modes are plotted against $\Delta t/\Delta x_1$ for $k_p\Delta x_1=0.2$. When the fastest growing modes are filtered out in a simulation, if the grid size is restricted to resolve the characteristic length of physical modes, the position of the $(\mu,\nu_1)=(0,0)$ mode can be moved to larger $\hat k_1$ by simply using a smaller time step.

Meanwhile, when the time step is fixed, the growth rates of higher order NCI ($\vert\nu_1\vert>1$) unstable modes can be efficiently reduced by using higher order particle shapes. In Fig. \ref{fig:allmodesgr} (a) we show how using different particle shapes changes the growth rate of the various NCI modes.  The parameters in Table \ref{tab:simpara} are used for this figure. The result indicates that, while using higher order particle shapes is very efficient in reducing the growth rate of higher order NCI modes, it is less efficient for the $(\mu,\nu_1)=(0,0)$ mode.  We also compared results with different grid sizes (while fixing $\Delta t/\Delta x_1=0.4$), as shown in Fig. \ref{fig:allmodesgr} (b). It indicates that reducing the grid size (while fixing $\Delta t/\Delta x_1$) helps reduce the growth rate of the $(\mu,\nu_1)=(0,0)$ mode, but not for the modes with $\nu_1\ne 0$.

\section{Strategies for eliminating NCI and sample simulations}
\label{sect:sampsim}
Based on this new understanding of the behavior of the unstable NCI modes, we now discuss approaches for controlling it. Once the NCI is adequately controlled, high fidelity simulations of relativistically drifting plasma can be carried out. We concentrate on spectral solvers and note that others are developing approaches for finite difference solvers \cite{GodfreyFDTDnotes}. The new form for the dispersion relation in Eq. (\ref{eq:2dmodesres}) can also be used to investigate the NCI for other solvers and we leave this for future work.

The approach we use is to first move the unstable modes to large $\bm{\hat{k}}$'s that are outside the region $\bm{\hat k}$ where important physics is occurring. As discussed in \cite{XuArxiv}, for the spectral solver the fastest growing modes at $(\mu,\nu_1)=(0,\pm 1)$ exist at large $\vert \bm{\hat{k}}\vert$ (the edge of the fundamental Brillouin zone). In addition, as discussed earlier in section \ref{sect:2d} their location in $\bm{\hat{k}}$-space does not change much as the grid sizes (for square or cubic cells) and time step are varied.

As discussed in section \ref{sect:2d} and shown in Fig. \ref{fig:variousnu1}, the second fastest growing mode at $(\mu,\nu_1)=(0,0)$ is highly localized in $\bm{\hat{k}}$-space and can be removed through a mask filter. However, these modes may exist near modes of physical interest, and for LWFA boosted frame simulations the plasma only exists in a small region of the simulation window. For such situations simply applying a mask filter may also effect the physics. We therefore eliminate those modes by first reducing the time step (while keeping the cell size fixed). As shown in Figs. \ref{fig:analytical} (a) and (b), this both moves the unstable modes to higher $\hat{k}_1$ and lowers the growth rate.

\begin{figure}[th]
\begin{center}\includegraphics[width=1\textwidth]{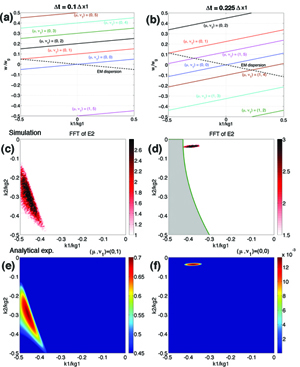}
\caption{\label{fig:variousnu1dt} (a) and (b) show the EM dispersion relation together with the beam resonance $\omega'-k'_1\beta=0$, for $\Delta t=0.1\Delta x_1$ and $\Delta t=0.225\Delta x_1$ (and other parameters the same as listed in Table \ref{tab:simpara}). (c) and (d) are the FFT of $E_2$ in the corresponding 2D simulations. The filter applied in order to observe the $(\mu,\nu_1)=(0,0)$ mode is illustrated by the grey areas in (d). (e) and (f) are the corresponding analytical predictions by using the expression Eq. (\ref{eq:asym1}) and (\ref{eq:asym2}).}
\end{center}
\end{figure}

To investigate how reducing the time step changes the NCI, 2D simulations using the same parameters as the those shown in Fig. \ref{fig:variousnu1}, but with a reduced time step of $\Delta t=0.1\Delta x_1$ are conducted. The corresponding beam resonances for this time step are illustrated in Fig. \ref{fig:variousnu1dt} (a), while the corresponding simulation data and analytical prediction for $\Delta t=0.1\Delta x_1$ are shown in Fig. \ref{fig:variousnu1dt} (c)--(f). From Fig. \ref{fig:variousnu1dt} (c) and (e) we see as expected that when the time step is reduced, the growth rate and pattern of the fastest growing modes at $(\mu,\nu_1)=(0,\pm 1)$ do not change much [compared with Fig. \ref{fig:variousnu1} (a) and (b)]. However, for the $(\mu,\nu_1)=(0,0)$ modes shown in Fig. \ref{fig:variousnu1dt} (d) and (f) , the locations move away from the center [compared with Fig. \ref{fig:variousnu1} (c) and (d)], while the growth rate is reduced by approximately a factor of 4.

In addition, when the time step is reduced to suppress the $(\mu,\nu_1)=(0,0)$ mode, the locations and growth rate for the higher order $\vert\nu_1\vert > 1$ modes also change. As seen in Fig. \ref{fig:variousnu1dt} (a), the next aliasing beam resonance after $(\mu,\nu_1)=(0,\pm 1)$ is $(\mu,\nu_1)=(0,\pm 2)$ rather than $(\mu,\nu_1)=(1,\pm 1)$. It is easy to see that in this case the $(\mu,\nu_1)=(0,\pm 2)$ resonance line has no intersection with the EM mode in the fundamental Brillouin zone. The beam resonance line for an intermediate time step of $\Delta t=0.225\Delta x_1$ is likewise shown in Fig. \ref{fig:variousnu1dt} (b). This illustrates how gradually reducing the time step changes the NCI modes in the fundamental zone.

Reducing the time step is preferable in relativistically drifting plasma simulation as it not only provides better NCI properties, but also provides better accuracy to Maxwell solver, and pusher in the algorithm. However, it comes at a cost of increased computational loads. In the following we describe another approach of eliminating the $(\mu,\nu_1)=(0,0)$ NCI modes for the spectral solver with a minor modification in the EM dispersion curve. This approach can be used alone, or combined with the reduced time step to achieve complete elimination of the $(\mu,\nu_1)=(0,0)$ NCI modes.

As seen from Fig. \ref{fig:rooteq}, the $(\mu,\nu_1)=(0,0)$ NCI modes are due to the intersection between EM mode and the main Langmuir modes at localized region in $\vec k$ space. To eliminate this intersection region, we now artificially create a small bump to the EM mode by slightly modifying the corresponding $[k]_1$ operator in the Maxwell solver
\begin{align}
[k]_1=k_1+\Delta k_{mod}
\end{align}
where
\begin{align}
\Delta k_{mod}=\Delta k_{mod,\max}\cos^2\biggl(\frac{k_1-k_{1m}}{k_{1,\min}-k_{1,\max}}\frac{\pi}{2}\biggr)\cos^2\biggl(\frac{k_2}{k_{2,\max}}\frac{\pi}{2}\biggr)
\end{align}
in the range $k_{1,\min}<\vert k_1\vert <k_{1,\max}$, and $\Delta k_{mod}=0$ otherwise. $k_{1,\min}$, $k_{1,\max}$, and $k_{inc,\max}$ are determined by the $(\mu,\nu_1)=(0,0)$ NCI modes to be eliminated, and $k_{1m}=(k_{1,\min}+k_{1,\max})/2$.

\begin{figure}[th]
\begin{center}\includegraphics[width=1\textwidth]{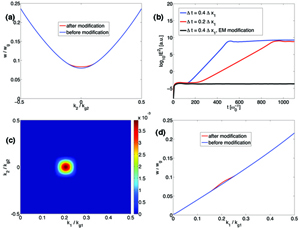}
\caption{\label{fig:bump} (a) shows the EM dispersion relation in vacuum before and after the modification, at line $\hat k_1=0.205$ and $\hat k_2 = 0$, while (c) shows the modification $\vert \omega-\omega'\vert/\omega_g$ in the fundamental Brillouin zone. (b) shows the $E_2$ energy evolution for simulations with $\Delta t=0.4\Delta x_1$ (with and without modification), and $\Delta t=0.2\Delta x_1$. Other simulation parameters are listed in Table \ref{tab:simpara}.}
\end{center}
\end{figure}

Consider the drifting plasma simulation discussed in section \ref{sect:2d} with $\Delta t=0.4\Delta x_1$ (and other simulation parameters are listed in Table \ref{tab:simpara}) as an example. In Fig. \ref{fig:bump} (a), (c), and (d) we illustrate how the EM dispersion (in vacuum) would change as we apply this modification to the $[k]_1$ operator in the solver in order to eliminate the $(\mu,\nu_1)=(0,0)$ NCI modes completely. In Fig. \ref{fig:bump} (c) we show the distribution of $\vert \omega-\omega'\vert/\omega_g$ to indicate how the EM dispersion is modified in the fundamental Brillouin zone, where $\omega$ and $\omega'$ are the frequency corresponding to a particular $(k_1,k_2)$ in the original, and revised EM dispersion, respectively. In Fig. \ref{fig:bump} (a) and (d) we show the corresponding EM dispersion for $(\omega,\hat k_2)$ and $(\omega',\hat k_2)$ at $\hat k_1=0.205$, and $(\omega,\hat k_1)$ and $(\omega',\hat k_1)$ at $\hat k_2 = 0$ respectively (as the lines of $\hat k_1=0.205$ and $\hat k_2 = 0$ cross the point where the maximum value of $\Delta k_{mod}$ is reached) to show how much the dispersion is modified. When substituting this $[k]_1$ operator in Eqs. (\ref{eq:asym1}) and (\ref{eq:asym2gen}) while keeping $[k]_2=k_2$, we can see there are is unstable root for $(\mu,\nu_1)=(0,0)$, i.e. when the modified $[k]_1$ operator is used in the solver, there is no $(\mu,\nu_1)=(0,0)$ NCI mode predicted by the theory. In this case $k_{1,\min}/k_{g1}=0.15$, $k_{1,\max}/k_{g1}=0.26$, $k_{2,\max}/k_{g2}=0.125$, and $k_{inc,\max}/k_{g1}=0.0095$. In Fig. \ref{fig:bump} (b) we plot the growth in energy for $E_2$ for the cases with $\Delta t=0.4\Delta x_1$ and $\Delta t=0.2\Delta x_1$, as well as the case with $\Delta t=0.4\Delta x_1$ plus the EM dispersion relation modification. In all these cases a low-pass filter is used to eliminate the fastest growing $(\mu,\nu_1)=(0,\pm 1)$ modes. As shown in Fig. \ref{fig:bump} (b) for the blue ($\Delta t=0.4\Delta x_1$), and red ($\Delta t=0.2\Delta x_1$) curve, the exponential energy growth is due to the $(\mu,\nu_1)=(0,0)$ modes; meanwhile in the case where the EM dispersion modification is applied (black curve), the energy growth due to $(\mu,\nu_1)=(0,0)$ modes is completely eliminated. Note later in time the energy grows exponentially (with a much lower growth rate, not shown in the plot) due to the higher order modes $(\mu,\nu_1)=(\pm 1, \pm 2)$. In these simulations we used second order particle shape. As discussed earlier in section 2, if one needs to further suppress the NCI by reducing the growth rate of the $(\mu,\nu_1)=(\pm 1, \pm 2)$ NCI modes, one can use a higher order particle shape as discussed in section.

In summary, we can first move the $(\mu,\nu_1)=(0,0)$ NCI modes away from physical modes by reducing the time step, then eliminate them by either applying a filter, or slightly modifying the EM dispersion in the highly localized region where the $(\mu,\nu_1)=(0,0)$ modes reside. One can take advantage of all these strategies available and combine them to obtain the best recipe for a particular application.

In the following, we will use these approaches to essentially eliminate the NCI in relativistic collisionless shock simulation, and in LWFA simulations in a Lorentz boosted frame, both of which involves the modeling of relativistically drifting plasma. In each case below we use second order particles.

\subsection{Relativistic collisionless shock}
\label{sect:sampsimshock}
\begin{figure}[th]
\begin{center}\includegraphics[width=1\textwidth]{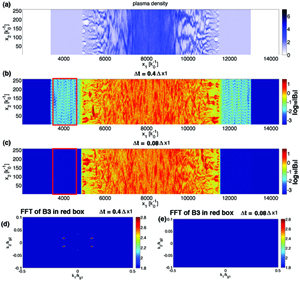}
\caption{\label{fig:shock} (a) shows the plasma density plot at $t =3360~\omega^{-1}_p$ for the $\Delta t = 0.08\Delta x_1$ case; (b) and (c) show the corresponding $\log_{10}\vert B_3\vert$ for the case $\Delta t = 0.4\Delta x_1$ and $\Delta t = 0.08\Delta x_1$, respectively. (d) and (e) shows the FFT of $B_3$ in the red box regions in (b) and (c), respectively. }
\end{center}
\end{figure}

In Fig. \ref{fig:shock} we present the results of two colliding plasma simulations, using the parameters in Table \ref{tab:shock}, with two different time steps. In these simulations we model the interaction of two counter-streaming plasma flows, each moving with a relativistic Lorentz factor of 20.0. Each plasma is initialized with a momentum distribution given by
\begin{align}
f(\bm p) \sim \exp\biggl(-\frac{(p_1-p_{10})^2}{2p^2_{th,1}}\biggr)\exp\biggl(-\frac{p^2_2}{2p^2_{th,2}}\biggr)\exp\biggl(-\frac{p^2_3}{2p^2_{th,3}}\biggr)
\end{align}
where $p_{10}$ and $\bm p_{th}$ are listed in Table \ref{tab:shock}. As the two flows interpenetrate they give rise to the so-called Weibel instability \cite{weibel}, which slows down the flows and forms two shocks that propagate in opposite directions. In both cases we use the low-pass filter to eliminate the $(\mu,\nu_1)=(0,\pm 1)$ NCI. Comparing the $\log_{10}\vert B_3\vert$ plots in Fig. \ref{fig:shock} (b) with $\Delta t=0.4\Delta x_1$ and (c) with $\Delta t=0.08\Delta x_1$, it is evident that when the time step is reduced, the noise originating from the NCI in the region where the two streams have not yet collided (overlap) with each other [shown in the red boxes in Fig. \ref{fig:shock} (b) and (c)] is much smaller. In Fig. \ref{fig:shock} (d) and (e) we also plot the FFT of the $B_3$ field for these same areas. The characteristic four-dot pattern of the $(\mu,\nu_1)=(0,0)$ modes is clearly observed only for $\Delta t=0.4\Delta x_1$. This illustrates that the $(\mu,\nu_1)=(0,0)$ modes can limit the length of the plasma that can be simulated even if the fastest growing modes are filtered out, and that these modes can be controlled by reducing the time step. The plasma density for the smaller time step at the same physical time is shown in Fig. \ref{fig:shock} (a) to show that there is no instability in the parts of the two streams that have not overlapped yet.

\begin{table}[t]
\centering
\begin{tabular}{p{6cm}c}
\hline\hline
\textbf{Parameters} & \textbf{Values}\\
\hline
grid size $(\Delta x_{1},\Delta x_{2})$ & $(0.5k^{-1}_p, 0.5k^{-1}_p)$\\
time step $\Delta t$ & $0.4\Delta x_1$, $0.08\Delta x_1$\\
number of grid & $32768\times 512$ \\
particle shape & quadratic\\
electron drifting momentum $p_{10}$ &19.975 $m_ec$\\
electron $\bm p_{th}$ & (0.001,0.001,0.001) $m_ec$\\
Ion mass ratio $m_i/m_e$ & 32\\
\hline\hline
\end{tabular}
\caption{Simulation parameters for the 2D shock simulation. $n_p$ is the plasma density, and $\omega^2_p=4\pi q^2n_p/m_e$, $k_p=\omega_p$ ($c$ is normalized to 1).}
\label{tab:shock}
\end{table}

\subsection{LWFA simulation in the Lorentz boosted frame}
\label{sect:sampsimlwfa}
We next present results from an LWFA boosted frame simulations in a nonlinear regime. The nonlinear regime is more challenging to simulate in the boosted frame due to self-trapping and the presence of wave harmonics. In Ref. \cite{YuArxiv} we showed excellent agreement between lab frame and boosted frame simulations are obtained in the linear regime using UPIC-EMMA when the fastest growing mode is filtered out. For simulation of nonlinear cases slight differences appear at higher $\gamma_b$. We revisit these simulations using strategies to systematically suppress the NCI modes.

Before we present the results, we note that in LWFA simulations the plasma density is not really a free parameter when the simulation is done in the wakefield frame where $\gamma_b=\gamma_w\equiv \omega_0/\omega_{p0}$, $\omega_0$ is the laser frequency. In this frame $\omega'_0=\gamma_b(\omega_0-k_0v_b)=2\omega_0/\gamma_b$, and $\omega^2_p/\gamma_b=\omega^2_{p0}$ is an invariant, which leads to
\begin{align}
\frac{\omega^2_p}{\gamma_b\omega'^2_0}=\frac{\omega^2_{p0}\gamma^2_b}{4\omega^2_0}=\frac{1}{4}
\end{align}
Therefore, with respect to $\omega'_0$ the value of $\omega^2_p/\gamma_b$ is fixed. The time steps and cell sizes are determined with respect to $\omega'_0$, therefore $\omega^2_p/\gamma_b$ is not a free parameter.

\begin{figure}[th]
\begin{center}\includegraphics[width=1\textwidth]{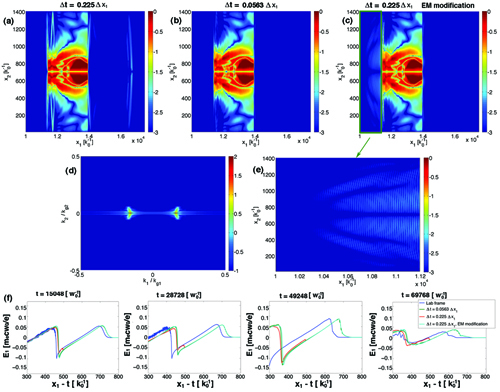}
\caption{\label{fig:lwfa} (a)--(c) shows the $\log_{10}\vert E_2\vert$ at $t=11135~\omega^{-1}_0$ for the cases with $\Delta t=0.225\Delta x_1$ (with and without EM modification), and $\Delta t=0.0563\Delta x_1$, respectively. In (e) we magnify the region in the green box in (c) to show the detailed structure of the radiation ahead of the drifting plasma, and (d) shows the corresponding FFT spectral for these radiation. (f) shows the on-axis $E_1$ wakefield when transforming the UPIC-EMMA simulations data back to the lab frame and compared again OSIRIS lab frame simulation data.}
\end{center}
\end{figure}

In Fig. \ref{fig:lwfa} we present results using parameters listed in Table \ref{tab:lwfapara1}.  These parameters are the same as in Ref. \cite{YuArxiv} with $\gamma_b=28$. The reference run used the time step $\Delta t=0.225\Delta x_1$, and additional cases were simulated to eliminate the NCI growth: a case with a reduced time step of $\Delta t=0.0563\Delta x_1$, and a case with $\Delta t=0.225\Delta x_1$ plus the EM dispersion modification (with the modification parameter $k_{1,\min}/k_{g1}=0.151$, $k_{1,\max}/k_{g1}=0.222$, $k_{2,\max}/k_{g2}=0.125$, and $k_{inc,\max}/k_{g1}=0.01$). The spatial resolution and number of simulation particles were kept fixed. In each case the low-pass filter is applied to eliminate the fastest growing NCI modes. In Figs. \ref{fig:lwfa} (a)--(c) we show the $\log_{10}\vert E_2\vert$ for the three cases at $t=11135~\omega^{-1}_0$. As is shown in Fig. \ref{fig:lwfa}, the self-injected particles observed in the case of $\Delta t=0.225\Delta x_1$ without EM dispersion modification [Fig. \ref{fig:lwfa} (a)] are no longer observable in the case with reduced time step $\Delta t=0.0563\Delta x_1$ [Fig. \ref{fig:lwfa} (b)], or the $\Delta t=0.225\Delta x_1$ case with EM dispersion modification [Fig. \ref{fig:lwfa} (c)]. Note in the 2D OSIRIS lab frame simulation, no self-injection particles are observed.

The fact that the $\Delta t=0.225\Delta x_1$ cases without EM dispersion modification shows self-injection particles, while the $\Delta t=0.225\Delta x_1$ case with dispersion modification and the $\Delta t=0.225\Delta x_1$ case there is no self-injection particle strongly indicates the $(\mu,\nu_1)=(0,0)$ NCI modes are interfering with the modeling of self-injection process. Slightly modifying the EM dispersion curve does not change the accuracy of the other parts of the algorithm (e.g. Maxwell solver, pusher), therefore the only difference between the two cases is in that for the modified-dispersion case there is no $(\mu,\nu_1)=(0,0)$ NCI modes, and the absence of these unphysical modes brings the simulation results closer to the lab frame results.

As a side note, we see in the green box in Fig. \ref{fig:lwfa} (c) there is radiation that is not seen in Fig. \ref{fig:lwfa} (a). This is due to the fact that when we artificially create a bump in the EM dispersion relation, part of the $\vec k$ in the bump has a group velocity difference to the drifting velocity of the plasma larger than the $\vec k$ outside the bump. As a result the radiation that is in the range of these $\vec k$ will travel faster than the other $\vec k$. We isolated the green box region and performed an FFT for the data inside the box, as shown in Figs. \ref{fig:lwfa} (d) and (e). We can see that the range of $\vec k$ for the radiation that is in the front of the drifting plasma corresponds exactly to those that has a larger group velocity.

The fact that both strategies bring the boosted frame simulation results closer to the lab frame results can also be seen by transforming the on-axis wakefields $E_1$ back to the lab frame and comparing them with lab frame OSIRIS simulation. In Fig. \ref{fig:lwfa} (f) we plot a lab frame time sequence of lineups of the on-axis wakefield. These plots correspond to the same time sequence as the second row of  Fig. 8 in Ref. \cite{YuArxiv}, and include some of the same data. Here we plot the line outs for the time steps $\Delta t=0.225\Delta x_1$ without the EM dispersion modification (red curve), $\Delta t=0.0563\Delta x_1$ (green curve), and $\Delta t=0.225\Delta x_1$ with the EM dispersion modification (cyan curve), and OSIRIS lab frame data (blue curve). It shows that better agreement with the lab frame result is found for the reduced time step, and for the case with larger time step plus EM dispersion modification. For the larger time step case without the EM dispersion modification (red curve) one can see the wake is perturbed at early times before the electric field reaches its minimum value in the rear of the first bubble. This is due to the self-trapped particles which are absent for the lab frame, and the two boosted frame simulations with elimination strategies applied to eliminate the $(\mu,\nu_1)=(0,0)$ NCI modes.

\begin{table}[t]
\centering
\begin{tabular}{lr}
\hline\hline
Plasma &\\
\quad density $n_0$& $1.148\times 10^{-3} n_0\gamma_b$\\
\quad length $L$ & $7.07\times 10^4k^{-1}_0/\gamma_b$\\
Laser & \\
\quad pulse length $\tau$ & $ 70.64k^{-1}_0\gamma_b(1+\beta_b)$\\
\quad pulse waist $W$ & $117.81k^{-1}_0$\\
\quad polarization & $\hat 3$-direction\\
2D boosted frame simulation&\\
\quad grid size $\Delta x_{1,2}$ & $0.0982k^{-1}_0\gamma_b(1+\beta_b)$\\
\quad time step $\Delta t/\Delta x_1$ & 0.225, 0.0563\\
\quad number of grid $(\gamma_b=28)$& 8192$\times$256\\
\quad particle shape & quadratic\\
\hline\hline
\end{tabular}
\caption{Parameters for the 2D LWFA simulations, with $a_0=4.0$. The laser frequency $\omega_0$ and laser wave number $k_0$ are used to normalize simulation parameters, and $n_0=m_e\omega^2_0/(4\pi e^2)$.}
\label{tab:lwfapara1}
\end{table}

\section{3D Scenario}
\label{sect:3dinst}
We next discuss the NCI in three dimensions. Based on the results for the 2D case, we write the full dispersion relation into the coupling between a Langmuir and an EM mode.
For the spectral solver, the dispersion relation for a specific  $\mu$, $\nu_1$ mode can be rewritten as,
\begin{align}\label{eq:3dmodesresspectralall}
&\left( [\omega]^2 - k_1^2  - k_2^2 -k^2_3 - \frac{\omega_p^2}{\gamma}(-1)^\mu S_j \frac{S_E[\omega] - S_B k_1 v_0}{\omega' - k'_1 v_0} \right)\nonumber\\
&\biggl\{\left((\omega' - k'_1 v_0)^2 - \frac{\omega_p^2}{\gamma^3}(-1)^\mu \frac{S_j S_E\omega'}{[\omega]} \right)\left( [\omega]^2 - k_1^2  - k_2^2 -k^2_3 - \frac{\omega_p^2}{\gamma}(-1)^\mu S_j \frac{S_E[\omega] - S_B k_1 v_0}{\omega' - k'_1 v_0} \right)  \nonumber \\
&+ \frac{\omega_p^2}{\gamma[\omega]} (-1)^\mu S_j(k_2^2+k^2_3) \{v_0^2(S_E\omega' - S_B[\omega])-v_0\nu_1S_Ek_{g1}\}\biggr\}=0
\end{align}
For the instability mode near the resonance line, we can assume
\begin{align}
&\left( [\omega]^2 - k_1^2  - k_2^2 -k^2_3 - \frac{\omega_p^2}{\gamma}(-1)^\mu S_j \frac{S_E[\omega] - S_B k_1 v_0}{\omega' - k'_1 v_0} \right)\ne0\\
&\label{eq:3dmodesresspectral}\left((\omega' - k'_1 v_0)^2 - \frac{\omega_p^2}{\gamma^3}(-1)^\mu \frac{S_j S_E\omega'}{[\omega]} \right)\left( [\omega]^2 - k_1^2  - k_2^2 -k^2_3 - \frac{\omega_p^2}{\gamma}(-1)^\mu S_j \frac{S_E[\omega] - S_B k_1 v_0}{\omega' - k'_1 v_0} \right)  \nonumber \\
&+ \frac{\omega_p^2}{\gamma[\omega]} (-1)^\mu S_j(k_2^2+k^2_3) \{v_0^2(S_E\omega' - S_B[\omega])-v_0\nu_1S_Ek_{g1}\}=0
\end{align}
which, as in the 2D case, can be viewed as the coupling between the Langmuir and EM mode. When Eq. (\ref{eq:3dmodesresspectral}) is compared with Eq. (\ref{eq:2dmodesresspectral}), we can see the equations in 3D can be obtained by replacing $k^2_2$ with $k^2_2+k^2_3$ in its 2D counterpart. As a result, the pattern of instability in 3D can be conveniently deduced. The location of the $(\mu, \nu_1)=(0,0)$ NCI modes in $\bm k$ space in 3D can be obtained as follows. Pick a point in $(k_2,k_3)$ space, then the growth rate and location in $k_1$ space of this mode will be the same as for $k^{2D}_2=\sqrt{(k^{3D}_2)^2+(k^{3D}_3)^2}$, where $k^{3D}_{1,2,3}$ and $k^{2D}_{1,2}$ are the coordinates of the modes in the 3D and 2D scenario respectively (assuming $\Delta x^{3D}_1=\Delta x^{3D}_2=\Delta x^{3D}_3=\Delta x^{2D}_1=\Delta x^{2D}_2$ and $\Delta t^{3D}=\Delta t^{2D}$). This indicates that the unstable modes form ``ring'' pattern in the $( k_2, k_3)$ space at specific values of $\hat k_1$. In addition, the maximum growth rate $\tau$ of these modes has $\tau^{2D}\approx \tau^{3D}$ when $\Delta x^{3D}_1=\Delta x^{3D}_2=\Delta x^{3D}_3=\Delta x^{2D}_1=\Delta x^{2D}_2$ and $\Delta t^{3D}=\Delta t^{2D}$.

\begin{figure}[th]
\begin{center}\includegraphics[width=1\textwidth]{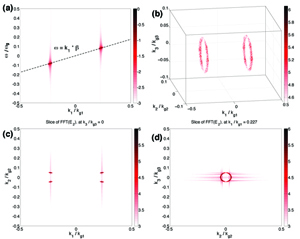}
\caption{\label{fig:inst3d} This figure shows the dominant NCI modes after the fastest growing modes are filtered out for a 3D simulation. (a) shows in 3D the $(\mu,\nu_1)=(0,0)$ mode which resides at the main resonance $\omega=k_1\beta$; (b), (c), and (d) are the positions of this NCI mode in $\bm{\hat k}$ space. }
\end{center}
\end{figure}

In Fig. \ref{fig:inst3d} we present data from a 3D simulation  of a drifting plasma, using the same parameters as in Table \ref{tab:simpara} except now $\Delta t^{3D}=0.35\Delta x_1$ (so that the Courant condition is satisfied). The different values of $\Delta t$ and the different noise sources in 3D v.s. 2D means the results will not be identical. We plot the FFT of $E_2$ in each panel. In Fig. \ref{fig:inst3d} (a) the real frequency v.s. $\hat{k}_1$ are plotted along with the line $\hat{\omega}_r=\hat{k}_1\beta$. This data was obtained for a line out along $x_1$ located at the middle of the box. In Fig. \ref{fig:inst3d} (b) a 3D plot of $E_2$ in $\bm{{k}}$-space is shown at a time during the exponential growth (before saturation). Only modes with amplitudes above 1/30 of the maximum mode are plotted. The predicted rings are clearly present. In Fig. \ref{fig:inst3d} (c) and (d) cross sections of the plot in Fig. \ref{fig:inst3d} (b) are shown.

In analogy with the 2D case, we can filter out this instability by applying a mask filter to eliminate the corresponding modes in the ring. For example we have a mask that blocks out all the modes between
\begin{align}
0.175\le \hat{k}_1 \le 0.275\qquad 0.027^2\le \hat{k}^2_2+\hat {k}^2_3 \le 0.067^2 \nonumber
\end{align}

\begin{figure}[th]
\begin{center}\includegraphics[width=1\textwidth]{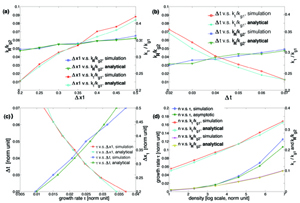}
\caption{\label{fig:tau_dt_dx_density_3d} Dependence of the position $(\hat k_1,\hat k_R)$, as well as the growth rate $\tau$ of the NCI at the fundamental Langmuir mode to grid sizes $\Delta x_1$ (with $\Delta x_1=\Delta x_2$ fixed), time step $\Delta t$, and plasma density $n_0$ in 3D. The variable $k_R$ refers to the radius of the ring pattern of $(\mu,\nu_1)=(0,0)$ NCI modes in 3D.}
\end{center}
\end{figure}

A parameter scan of the growth rate and position of the unstable modes using UPIC-EMMA, as well as comparison between the analytical predictions is presented in Fig. \ref{fig:tau_dt_dx_density_3d} (a)--(d). The variable $k_R$ in these plots refers to the radius of the ring pattern. It confirms that when comparing the unstable mode for a 2D case (see Fig. \ref{fig:tau_dt_dx_density}) against its counterpart in 3D, we have approximately $\hat k^{3D}_{1}=\hat k^{2D}_{1}$, and $\hat k^2_R\equiv(\hat k^{3D}_{2})^2+(\hat k^{3D}_{3})^2=(\hat k^{2D}_{2})^2$ (where $\hat k_R=k_R/k_{g1}$), and $\tau^{3D}\approx\tau^{2D}$. Therefore, the 3D NCI can effectively be eliminated by using the same strategies as in 2D.

\begin{figure}[p]
\begin{center}
\includegraphics[width=1.04\textwidth]{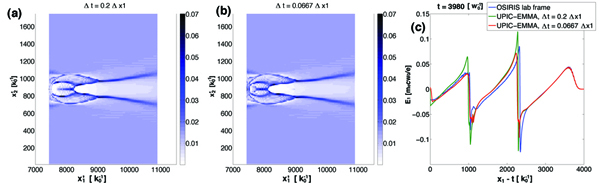}
\caption{Results from 3D UPIC-EMMA boosted frame simulation ($\gamma=17$). (a) and (b) present 2D cross section plots of the plasma electron density for $\Delta t=0.2\Delta x_1$ and $\Delta t=0.0667\Delta x_1$, while (c) shows the on-axis $E_1$ comparison between the three cases at $t=3980~\omega^{-1}_0$ in the lab frame. $x_1-t$ is the coordinates moving together with the moving window.}
\label{fig:lu3d}
\end{center}
\end{figure}

As an example of a three dimensional case,  we revisit the 3D LWFA boosted frame simulation presented in Ref. \cite{YuArxiv} using a time step of $\Delta t=0.2\Delta x_1$ where only a low pass filter was used to eliminate the $(0,\pm 1)$ mode. Results are shown in Fig. \ref{fig:lu3d}. The simulation was rerun with the reduced time step of $\Delta t=0.0667\Delta x_1$ to ensure elimination of the $(0,0)$ modes  while  a low-pass filter is used to eliminate the $(0,\pm 1)$ mode. Other parameters for these two simulations are listed in Table 2 of Ref. \cite{YuArxiv}. As seen in Fig. \ref{fig:lu3d}, the plasma density is similar for the two time steps of the two boosted frame cases, although there are subtle differences.  In  Fig. \ref{fig:lu3d} (c), data is transformed back to the lab frame and a line out of the $E_1$ is plotted.  There is general agreement between all three cases. Nonetheless, it is interesting to note that there are differences. As seen in Fig. \ref{fig:lu3d} (c) the $\Delta t=0.0667\Delta x_1$ case (red curve) shows slightly heavier beam loading in the first bucket than in the $\Delta t=0.2\Delta x_1$ case (green curve), and both boosted frame cases show more beam loading than in the lab frame case (blue curve). There are also differences in the later buckets. This result, and the 2D results discussed in section \ref{sect:sampsimlwfa} reveals the fact that the modeling nonlinear wakes and the self-injection process is much more challenging than modeling linear or weakly nonlinear wake fields (where there is generally better agreement). These results also show that in any given simulation it is not obvious if the higher order NCI modes are an issue. However, as alluded to earlier, the spectral (FFT based) solvers have the advantage that the accuracy of the simulation for given cell sizes and number of particles per cell improve as the time step is reduced. This occurs because the NCI modes are less of an issue and the overall accuracy of the particle push and numerical dispersion also improve.

\section{Summary}
\label{sect:summary}
We have systematically investigate the unstable NCI modes with an emphasis for a spectral (FFT) based Maxwell solver. We start from a general dispersion relation described in previous work \cite{XuArxiv}. This previous work and that of others shows that the unstable NCI modes occur near the intersection of wave particle resonances (including aliases), i.e., where $\omega'-k'_1\beta=0$, and electromagnetic waves. Therefore, each unstable mode can be identified as coming from a specific value of $\mu$ and $\nu_1$. Based on this information, we rewrite the dispersion relation in both 2D and 3D for given values of $\mu$ and $\nu_1$ into the coupling between two modes which are each numerically stable, and for which the coupling term vanishes in the continuous limit. These two modes are easily identified as the Lorentz transformed version of modes which are purely longitudinal (Langmuir) and purely transverse (EM) in the rest frame of the plasma. We then use the new form of the NCI dispersion relation to study the NCI for a spectral (FFT) based Maxwell solver. The fastest growing modes corresponding to $(\mu,\nu_1)=(0,\pm 1)$ were studied in Ref. \cite{XuArxiv} and \cite{YuArxiv}. Here, we studied additional modes with $(\mu,\nu_1)=(0,0)$ and $\vert \nu_1\vert>1$. We find for the FFT solver that the second fastest growing mode is at $(\mu,\nu_1)=(0,0)$ and that unlike the $(\mu,\nu_1)=(0,\pm 1)$ mode its location moves to larger values of $\hat k_1$ and its growth rate reduces as $\Delta t/\Delta x_1$ decreases.

Unlike the NCI modes with $\vert\nu_1\vert>0$, the NCI at the fundamental Langmuir resonance has a highly localized pattern of four dots in the $\bm k$ space in 2D (one in each quadrant); while in 3D  the unstable modes form a ring pattern. A simple relation of the growth rate and locations of this instability between the 2D scenario and 3D scenario is found with $\hat k^{3D}_{1}=\hat k^{2D}_{1}$, and $(\hat k^{3D}_{2})^2+(\hat k^{3D}_{3})^2=(\hat k^{2D}_{2})^2$, and $\tau^{2D}\approx \tau^{3D}$ when $\Delta x^{3D}_1=\Delta x^{3D}_2=\Delta x^{3D}_3=\Delta x^{2D}_1=\Delta x^{2D}_2$ and $\Delta t^{3D}=\Delta t^{2D}$ is assumed.

Based on the new understanding of the family of unstable NCI modes, we developed strategies for eliminating them with an emphasis for a spectral solver. The principle idea is to ensure the unstable modes are far away from the physics of interest and to reduce their growth rate so that they do not grow during the simulation. Note that according to Eq. (\ref{eq:nu1position}) the $\nu_1=\pm 1$ NCI always resides far away from the interested physics. The $\nu_1=0$ mode can be moved away towards large $\bm{\hat k_1}$ by reducing the time step. In principle, a dedicated mask filter can then be applied in the solver to eliminate these modes without affecting the modes of physical interest. However, since the growth rate of the $\nu_1=0$ NCI decreases as the time step is decreased, it may not be necessary to apply the mask filter to eliminate this mode. In this paper, we proposed two methods for eliminating the $\nu_1=0$ modes. In one we reduce the time step, while in the other we slightly modify the EM dispersion curve near the region in $\vec k$ space where there is coupling between the EM modes and main Langmuir mode. The growth rates of the higher order NCI modes with $\vert\nu_1\vert>1$ are reduced when using higher order particle shapes and they are also modified as the time step is reduced. We show that in UPIC-EMMA simulations of both LWFA  in a Lorentz boosted frame and of colliding plasmas (as is done when simulating relativistic collisionless shocks) can be carried out with no evidence of the NCI. In both cases the results show that the $(\mu,\nu_1)=(0,\pm 1)$ NCI modes are effectively eliminated when a low pass filter is used, and the $(\mu,\nu_1)=(0,0)$ modes are elimianted when using a reduced time step or applying EM dispersion modification in the Maxwell solver.

For future work we will investigate the tradeoff in the mitigation strategies discussed in this paper. When doing a physics simulation signals at all values of $\bm{\hat k}$ occur. So determining \textit{a priori} what time step can be used is difficult. However, one advantage of the spectral solver is the accuracy of all the physics improves as the time step is reduced, which is not true for finite difference solvers as this can lead to large dispersion errors in EM waves. So once a time step is found for which the results have converged, one can investigate how the results compare with a combination of modified EM dispersion and intermediate time step. Furthermore, studying how the physics changes as the time step is reduced and how results actually converge is also an important area for future work. Determining where the NCI will occur in time and space using the predicted growth rates and then mapping this to time and space in the lab frame is another area for future work. Lastly, it will be useful to develop filters and interpolation schemes that theoretically cancel the coupling term, $\mathcal{C}$. This approach would overlap with the finite difference solver investigations of \cite{GodfreyFDTDnotes}.

This work was supported by US DOE under grants DE-SC0008491, DE-SC0008316,  DE-FC02-04ER54789, DE-FG02-92ER40727, by the US National Science Foundation under the grant ACI 1339893, and by NSFC Grant 11175102, thousand young talents program, and by FCT (Portugal), grant EXPL/FIS-PLA/0834/1012, and by the European Research Council (ERC-2010-AdG Grant 267841), and by LLNL's Lawrence Fellowship. Simulations were carried out on the UCLA Hoffman2 and Dawson2 Clusters, and on Hopper cluster  of the National Energy Research Scientific Computing Center.


\end{document}